\documentstyle[sprocl]{article}

\input{psfig}

\newcommand{\nc}{\newcommand}
\nc{\lnc}{\Lambda_{NC}}
\nc{\tmn}{\theta_{\mu\nu}}
\nc{\tmf}{\theta_{\mu5}}
\nc{\th}{\theta}
\nc{\ps}{\psi}
\nc{\psl}{\slash{\!\!\!p}}
\nc{\fr}[2]{\frac{#1}{#2}}
\nc{\si}{\sigma}
\nc{\Tr}{\mbox{Tr}}
\nc{\bi}[1]{\bibitem{#1}}

\def\be{\begin{equation}}
\def\ee{\end{equation}}
\def\bea{\begin{eqnarray}}
\def\eea{\end{eqnarray}}
\begin{document}

\title{LIMITS ON THE NON-COMMUTATIVITY SCALE}
\author{I. MOCIOIU${}^*$, M.POSPELOV${}^+{}^\$ $ 
R. ROIBAN${}^*{}^\%$}
\address{${}^*$C.N.Yang ITP
 SUNY Stony Brook, NY 11794-3840\\
 E-mail:mocioiu, roiban@insti.physics.sunysb.edu\\
${}^+$Theoretical Physics Institute, School of Physics 
and Astronomy, 
 University of Minnesota, 116 Church St., Minneapolis, MN 55455
 E-mail:pospelov@mnhepw.hep.umn.edu\\
${}^\$ $ McGill University, 3600 University Street, Montreal, 
Quebec, Canada H3A 2T8\\
${}^\% $Physics Department, 
University of California, Santa Barbara, CA, 93106-9530}

%\author{I. MOCIOIU${}^*$, M.POSPELOV${}^+$, R. ROIBAN${}^*$}

%\address{${}^*$C.N.Yang ITP
% SUNY Stony Brook, NY 11794-3840
%\\ E-mail:mocioiu, roiban@insti.physics.sunysb.edu
%}

%\address{${}^+$Theoretical Physics Institute, School of Physics 
%and Astronomy \\
% University of Minnesota, 116 Church St., Minneapolis, MN 55455
%\\ E-mail:pospelov@mnhepw.hep.umn.edu}

\maketitle\abstracts{
A non-vanishing vacuum expectation value for an antisymmetric tensor field 
leads to the violation of Lorentz invariance, controlled 
by the dimension ($-2$) parameter, $\theta _{\mu \nu }$.
We assume that the zeroth order term in $\theta $-expansion 
represents the Standard Model and study the effects induced by linear terms 
in $\theta _{\mu \nu }$. If coupling to $\theta_{\mu\nu}$ is realized in 
strongly interacting sector of the theory, the clock comparison 
experiments place the limit on 
the possible size of this background at the level of 
$1/\sqrt{\theta }\mathrel{\raise.3ex\hbox{$>$\kern-.75em\lower1ex%
\hbox{$\sim$}}}5~\times 10^{14}$ GeV. If the interaction with 
$\theta_{\mu \nu }$ is initially present only in the QED sector,
this limit can be relaxed to $10^{11}-10^{12}$ GeV level. 
The strength of these limits obviates the inferiority of collider
physics with regard to experimental checks of Lorentz invariance. 
Limits of similar strength are 
expected to hold in the case of mixed non-commutativity between 
four-dimensional and extra-dimensional coordinates. We also show that 
in certain models mixed non-commutativity can be interpreted as 
4d CPT-violating background. 
}

\section{Introduction}

Non-commutative field theories and their string theory realizations 
have been a subject of intensive theoretical research over the past few 
years (See, e.g.\cite{SWDN}). They can be defined either as theories on a 
space in which the coordinates are self-adjoint operators satisfying
$[{\hat x}^\mu,\,{\hat x}^\nu]=i\theta^{\mu\nu}$ and
$[\theta^{\mu\nu},\,{\hat x}^\rho]=0$ or as theories on an ordinary space
but with a deformed multiplication law. 
It is easy to see that the operator
$\theta^{-1}_{\mu\nu}[{\hat x}^\nu,\,.]$ defines 
a derivation of the product of function on this space. This observation allows 
one to write generic field theory actions. 
%Translation to the second
%approach is achieved by using the Weyl symbol defined as
Translation between the two approaches is done using the Weyl symbol
\be
{\hat \Phi}({\hat x})=\int d^D x
\int{d^dk\over (2\pi)^d} \phi(x) {\sl e}^{ik_\mu{\hat x}^\mu}
{\sl e}^{-ik_\mu{x}^\mu}~~.
\ee
If $V({\hat\Phi}({\hat x}))$ is an arbitrary function of fields 
${\hat \Phi}({\hat x})$ which we will interpret as potential, the
equivalent action in terms of $\phi(x)$ is 
\be
S=\int d^d x [{1\over 2}(\partial_{\mu}\phi)^2+
V_*(\phi) ]
\ee
where $V_*(\phi)$ is  $V(\hat\Phi)$ in which we replace $\hat\Phi$
by $\phi$ and multiplication is given by the $*$-product
\be
(\phi\star \psi)(x)=
e^{\,\frac{i}{2}\theta^{\mu\nu}\partial_\mu(x) \partial_\nu(y)}{\phi}(x)
{\psi}(y)|_{x=y}
\ee

One interesting phenomenological question is how large $\tmn$ could be without
contradicting existing experimental data. To answer this, we start with the 
Standard Model and we include the external background $\tmn$ via  the Moyal 
product. We take the (*)-modified~Standard Model and expand it once in the 
external $\theta $ parameter. The result is
that SM is extended by the series of dimension 6 operators, composed from
three or more fields: 
\begin{equation}
SM(*)=SM+\sum_{i}\theta _{\mu \nu }O_{\mu \nu }^{(i)}.  \label{sm*}
\end{equation}

We consider this as an effective theory with a cutoff smaller than 
$1/\sqrt{\theta}$. The higher the energy/momentum transfer is, the 
larger the effect of $O_{\mu \nu }$ will be, so one can think of 
obtaining stringent limits 
on $\theta $ by using high-energy processes.
We have considered in \cite{us1} the Z-boson decay into a pair of photons,
forbidden by Lorentz invariance and Bose statistics in the SM, but allowed
when  $\theta_{\mu \nu }\neq 0$.
We obtained a very modest limit, $1/\sqrt{\theta }> 250$GeV. This can be
slightly improved by considering $\theta $-induced corrections to other
high-energy processes \cite{HewHinch}, but the experiments do
not have enough accuracy to produce sufficiently strong bounds on 
$\theta _{\mu \nu }$, so we turn our attention to low energy experiments.
We show \cite{us1} that low-energy  precision experiments
place the limit on the possible size of $\theta$ at the level of 
$1/\sqrt{\theta }\mathrel{\raise.3ex\hbox{$>$\kern-.75em\lower1ex%
\hbox{$\sim$}}}5~\times 10^{14}$ GeV. 
% The presence of a mixed component of the 
% non-commutative parameter $\theta_{\mu M}$, where $\mu=0,1,2,3$ and M is 
% an extra dimensional index can also 
% lead \cite{us2} to violation of four-dimensional
% CPT invariance. 
% Starting from five-dimensional $U(1)$ theory we compute
% the radiatively-induced couplings of $\theta _{\mu 5 }$ with the 
% four-dimensional axial vector current and with the CPT odd operators 
% of dimension 6. The result is that  
% $1/\sqrt{\theta_{\mu 5 }}$ needs to be higher than
%  $5~\times 10^{11}$ GeV for the effects to be within
% the errors of clock comparison experiments 

\section{Experimental limits on $\theta_{\mu\nu}$}

The most notable feature of the effective Lagrangian (\ref{sm*}) is the
explicit violation of Lorentz invariance. Such a feature is 
experimentaly excluded to a very high accuracy 
(see, e.g. \cite{Kost}).  
This violation is controlled in our case by the size of $\theta_{\mu\nu}$
and its effect first shows up in dimension 6 operators. One can show 
\cite{us1}, that the non-commutative QED leads to the interaction 
of atomic angular momentum $J_i$ with the ``magnetic'' component 
$\epsilon_{ijk}\theta_{ij}$. 
At this point it becomes clear that similar effects
will appear in the hadronic physics, when we consider
the (*)-extended interaction of quarks and gluons. In the most interesting 
case, this interaction leads to
the effective coupling of the nucleon spin with $\theta $,
\begin{equation}
\langle N|{\cal L}_{QCD}(*)|N\rangle =~
{\rm \theta\!-\!independent~terms}~~+~~\frac{d_{\theta }}{2}\theta _{\mu \nu }
\overline{N}\sigma _{\mu \nu }N,
\label{generic}
\end{equation}
which in non-relativistic limit simply becomes  $d_{\theta }({\bf \theta }
_{B}\cdot \frac{{\bf S}}{S})$. The size of $d_{\theta }$ is now given by 
the third power of a characteristic hadronic scale, $ d_{\theta }\sim 
\Lambda_{\rm had}^3$. One can further discuss 
the issue \cite{loops}
whether $\theta_{\mu\nu} \bar q \sigma_{\mu\nu} q$ can be generated 
through loops at the quark level. In this case, one can hope to substitute  
two powers of $\Lambda_{\rm had}$ by the scale of the effective cutoff, 
so that $d_\theta \sim ({\rm loop ~ factors})^n \Lambda_{\rm had} 
\Lambda_{\rm cutoff}^2$, where $n$ is the number of loops involved. 
It is likely that $n=2$. Since the value of $\Lambda_{\rm cutoff}$ is 
quite uncertain and could be low, the loop level does not present serious 
advantage over $\Lambda_{\rm had}^3$ and we choose to limit our discussion to 
the tree level. 

To estimate $d_{\theta }$ we use a simplified version of the nucleon
three-point function QCD sum rules \cite{Nsr}. 
We compute the two point function
$
\Pi(p)=i\int d^4 x e^{ip\cdot x}\langle 0|\eta(0){\bar \eta}(x)|0\rangle
$
of generalized spin $1/2$ nucleon interpolating currents: 
$
\eta(x)=2\epsilon_{abc}\{\left[u^a{}^T(x)C\gamma_5d^b(x)\right]+
\beta\left[u^a{}^T(x)C d^b(x)\right]\gamma_5\} \,u^c(x)$
in the presence of the external $
\theta _{\mu \nu }$-background. $\beta$ is a free parameter.
% usually taken between -1 and 1 which can be used for 
% the optimization of sum rules. For simplicity, 
Here we use $\beta=0$ correlator 
that has the best overlap with the nucleon ground state. 
By quark-hadron duality,  the result must also be 
represented as a sum of resonance contributions which can be created 
by $\eta$ current from vacuum,$\langle 0|\eta|N\rangle = \lambda u_N$.
To extract the desired information about the ground state nucleon, we
need to be in the intermediate regime where $-p^2$ is high enough that 
asymptotically free calculations make sense, while $-p^2$ should be low 
enough such that on the ``phenomenological'' side only the ground state 
survives and the contribution of the excited states is suppressed.
We match the two sides at 1 GeV to obtain an
estimate of $d_{\theta }$ for nucleons.

On the OPE\ side we can use an asymptotically free description, and thus
include the $\theta _{\mu \nu }$-piece as the correction to a free massless
quark propagator:
\begin{equation}
S(x,0)=\frac{ix_{\mu }\gamma _{\mu }}{2\pi ^{2}x^{4}}-\frac{ix_{\mu }\gamma
_{\beta }}{2\pi ^{2}x^{4}}t^{a}G_{\nu \beta }^{a}\theta _{\mu \nu }.
\end{equation}
It is then straightforward to compute the two-point function
and express the result as the combination of a pure perturbative
piece  and non-perturbative condensates.
Choosing $\sigma_{\mu\nu}\theta_{\mu\nu}$ Lorentz channel,
performing Borel transformation and
equating OPE side to the phenomenological part we obtain:
\be
(\theta\cdot\sigma){1\over 108\pi^2}m_0^2\langle{\bar q}q\rangle +
{\cal O}(1/M^2)
=\lambda^2\,{{1\over 2}m_N^2 d_\theta\over M^4} e^{-m_N^2\over M^2}
(\theta\cdot\sigma)+{\cal O}(e^{-m_N'\over M^2})
\ee
with $m_0^2=\langle \overline{q}
\sigma\cdot G q\rangle/\langle \overline{q}q\rangle \simeq 0.8$ GeV$^2$, and 
$(2\pi)^4 \lambda^2\simeq 1$ GeV$^6$, 
from which we get our estimate of $d_\theta$:
\be
d_\theta\sim {1\over 54\pi^2}m_0^2\langle{\bar q}q\rangle
{M^6\over m_N^4}e^{m_N^2\over M^2}{1\over \lambda^2}\,\simeq \,0.1\,{\rm GeV}^3
\ee
This estimate can be improved by calculating next order terms in OPE, 
including the anomalous dimensions, 
and fixing the size of unknown ''susceptibility'' condensates that 
would necessarily arise in the presence of $\theta_{\mu\nu}$ background. 

The interaction of the form (\ref{generic}) creates a 
Zeeman-type splitting of atomic multiplets relative to 
the fixed vector $\theta_B$ and can be searched through the 
siderial variations of the Zeeman frequencies. Thus, one can use the results 
of the most precise clock comparison experiments \cite{Amh,Harvard1}. 
The presence of $\tmn$-background affects primarily nuclear spin. Thus the 
magnetic field, measured by mercury atom, will be corrected due to the 
interaction of the nuclear spin with $\theta_B$ to a larger extent than the 
magnetic field measured by paramagnetic Cs. 
The absence of sidereal variations 
in the difference of magnetic field measured by Cs and Hg verified 
\cite{Amh} at
the level of 100 nHz translates into the  extremely tight bound:
${1}/{\sqrt\theta} > 5\cdot 10^{14}~ {\rm GeV}
\label{limit}$. 
If non-commutativity is introduced initially only in the QED sector,
the interaction (\ref{generic})
is still generated, although suppressed by $(\alpha/\alpha_s)^2$ 
compared to the  QCD case. This lowers the limit to 
the level of  $\theta <
(10^{11}-10^{12}{\rm GeV})^{-2}$ which is still beyond the reach of any 
concievable high-energy experiment.

\section{Breaking CPT by mixed non-commutativity}

We now discuss the phenomenological consequences of mixed non-commutativity 
$\theta_{\mu M}$, where $\mu$ is a normal 4-d index and 
$M$ is along extra dimensions. An intriguing possible consequence  of 
this interaction is the breaking of the four-dimensional CPT symmetry. 
To understand the origin of this breaking, one can consider 
the effective loop-generated \cite{us2,loops} interaction between 
a fermion $\psi$ and 5d non-commutative background. 
$ \theta_{MN}\bar \psi \sigma_{MN} \psi$. Specifying  $M$ and $N$ 
to $\mu$ and $5$, we get 
$ \theta_{\mu 5 }\bar \psi \gamma_\mu\gamma_5 \psi$ which is explicitly 
CPT-odd in four dimensions.

In order to get a quantitative estimate of the magnitude 
of the effective CPT violation we consider a 
5-dimensional non-commutative theory with all fields 
propagating in all 5 dimensions, compactify the fifth dimension on a
circle and integrate out the heavy Kaluza-Klein modes to obtain the low 
energy effective action for the zero-modes. 
Computing one-loop diagrams in the 5-d non-commutative U(1) theory we obtain 
CPT non-invariant $dim=6$ operators in four dimensions:
\be
{\cal L}=\fr{5e_i^2e}{48\pi^2}\left[1+\fr{9}{10}\fr{e_i}{e}\right]
\theta_{\mu5}
(\bar\psi_i i\si^{\alpha\beta} \gamma_5\psi_i) \partial_\mu F_{\alpha\beta},
\label{4ferm}
\ee
which is independent of the mass of the heavy KK mode.
Phenomenological consequences of this interaction are obtained
by specializing (\ref{4ferm}) to the case of the light quarks.
The relevant matrix elements over nucleon and nuclear states 
turn out to be rather simple in this case. We refer 
a reader to our original paper \cite{us2} for more details. 
In the end, the effective operator (\ref{4ferm}) translates to the 
interaction of the form $I_i \theta_{i5}$, where $I$ is the nuclear spin. 
Using again the results of the two most sensitive 
experiments \cite{Amh,Harvard1}, 
we deduce the level of sensitivity to the presence of mixed non-commutativity
$
|\theta_{i5}|  \mathrel{\raise.3ex\hbox{
$<$\kern-.75em\lower1ex%
\hbox{$\sim$}}} ( 5\cdot 10^{11} {\rm GeV}    )^{-2}
\label{finlim}
$
It should be mentioned here that CPT-violating terms can be avoided in 
orbifold models or in models where fermions are localized to 
a 3-dimensional domain wall. 

\vspace{-2pt}

\section{Conclusions}

Extreme precision of clock comparison experiments impose severe 
restrictions on the possible scale of non-commutativity.
Non-commutative QCD requires 
${1}/{\sqrt{\theta_{ij} }}\mathrel{\raise.3ex\hbox{
$>$\kern-.75em\lower1ex%
\hbox{$\sim$}}}5\times 10^{14}$ GeV, non-commutative QED 
implies 
${1}/{\sqrt{\theta_{ij} }}\mathrel{\raise.3ex\hbox{
$>$\kern-.75em\lower1ex%
\hbox{$\sim$}}}10^{11}-10^{12}$ GeV
while the limits on mixed non-commutativity are 
${1}/{\sqrt{\theta_{\mu 5} }}\mathrel{\raise.3ex\hbox{
$>$\kern-.75em\lower1ex%
\hbox{$\sim$}}}5\times 10^{11}$~GeV.

\vspace{-3pt}
\section*{Acknowledgments}
The work of IM and RR was supported by NSF Grant PHY-9722101 and 
the work of MP was supported by DOE Grant DE-FG02-94ER40823. RR was 
also supported by NSF under Grant No. 0098395 and DOE under Grant 
No.91ER40618(3N).

\end{document}